\documentclass[prd,showpacs,nofootinbib,preprintnumbers]{revtex4}
\usepackage{latexsym,amsmath,amssymb,amsbsy,graphicx}

\renewcommand{\mathbf}{\boldsymbol}

\newcommand{\suma}{\sum_{\raisebox{2pt}{\smash{$\scriptstyle a$}}} }

\newcommand{\bs}{\beta^{{\kern 0.5pt}\prime}}

\newcommand{\ds}{\displaystyle}

\newcommand{\kn}{{\kern 1pt}}
\newcommand{\akn}{{\kern -1pt}}

\newcommand{\e}{\mathrm{e}}

\newcommand{\pp}{\, ... \,}
\newcommand{\nn}{\nonumber}

\newcommand{\ie}{{\scriptscriptstyle <}}
\newcommand{\et}{{\scriptscriptstyle >}}
\newcommand{\el}{ \Delta_{\rm \scriptscriptstyle E}}
\newcommand{\eu}{{\rm \scriptscriptstyle E}}
\newcommand{\ee}{\end{equation}}
\newcommand{\be}{\begin{equation}}

 \numberwithin{equation}{section}

\begin {document}
\title{Casimir interaction of finite-width strings}
\author{Yuri V. Grats\footnote{E-mail: grats@phys.msu.ru} and Pavel Spirin \footnote{E-mail: pspirin@physics.uoc.gr}}
\affiliation{ Department of Theoretical Physics, Moscow State
University,119991, Moscow, Russian Federation}

j
\begin{abstract}
Within the trln-formalism we investigate the  vacuum interaction of  cosmic strings and the influence of strings width on this effect.
For the massless real scalar field we compute the  Casimir contribution into the total vacuum energy. The dimensional-regularization technique is used.
It is shown that the regularized Casimir term contains neither the UV-divergences, nor the divergences related with the non-integrability of the renormalized vacuum mean of the energy-momentum tensor.

In the case  of two infinitely-thin strings, the limit   coincides with the known   result. The effect related with the finite width becomes significant on the distances of  several  core diameters.\\[7pt]
\end{abstract}

\maketitle


\section{Introduction}

Since the mid of XX century, the Casimir effect attracts the more attention.
Being direct evidence of the relationship between quantized fields and macroscopic external phenomena, it is of interest both from the fundamental physics viewpoint, and from the applied physics.
 Making change to the vacuum fluctuation spectra, the external conditions generate the finite additions to the vacuum energy. It influences to the observable forces acting on  macroscopic objects.
 The Casimir interaction  valuably affects on various processes on various space-time scales. Now it is an object of research not only by  physicists working in QFT, atomic physics,
 nanotechnology or condensed-matter physics, but also by scientists working in gravitation and cosmology
(see, e.g. \cite{Bordag}).
Among the problems of the latter type we can notify the problem of vacuum interaction of the cosmic strings.

Cosmic strings are one-dimensional extended (infinite or closed) topological defects, which might be generated under the cosmological phase transitions
\cite{Dowker,Vilenkin}. Observable data on the cosmic microwave background exclude cosmic strings from the set of basic sources of primary fluctuations of the Universe density. However, they still be considered as a possible reason of a number of observable effects (the review on the possible appearance of cosmic strings is available in
\cite{Copeland}). This stimulates a search of ways to detect the cosmic strings and, consequently, motivates the investigation of phenomena related with the behavior of classical/quantum matter in conical spaces.

Here we restrict ourselves by  consideration of the space-time generated by a system  of parallel straight infinitely long cosmic strings. The striking feature of straight cosmic string is a so called ''gravitational sterility'' of it. This consists, in particular, in the absence of gravitational interaction between the parallel strings.
However, the global distinction of conical spaces from the Minkowski one leads to the change of vacuum fluctuations of quantum fields and, consequently, to the appearance of the attraction force between strings.

The first estimate of this effect for infinitely thin strings was obtained in
\cite{Bordag1990}.
Subsequently this result was refined in a series of works \cite{Grats1995,Bordag2014,Grats2016}.

In the work \cite{Grats1995} we   considered the vacuum interaction of cosmic strings within the local one-loop approach, where the object of research was a vacuum expectation value of the operator of energy-momentum tensor.
It was shown that to the lowest order of perturbation theory, the contributions into
~$ \langle\kn T_{\mu\nu}\rangle_{\rm vac}^{\rm ren}$  come additively from different strings, while the Casimir contribution into the vacuum mean (which depends upon the inter-string distance) reveals itself in the second perturbational order.

 Meanwhile, the computation of total vacuum energy encounters the additional difficulty.
Namely, the vacuum energy is determined by the integration of energy density  $\langle T_{00}\rangle_{\rm vac}^{\rm ren}$ over whole space.
But in the case of infinitely thin string, the renormalized vacuum mean has a non-integrable singularity at the string location
 \cite{Frolov,Frolov1,Frolov2,Frolov3,Frolov4,Frolov5}. Hence the elimination of ultra-violet divergences does not fix the whole problem.  As it was shown in
\cite{Fursaev}, one may eliminate additional divergence by renormalization of the bare string tension.
In \cite{Grats1995} we took it into account and showed that if exclude from   ${\cal E}_{\rm vac}$
the terms insensitive to the inter-string distance\footnote{A standard rule, which is of usage when one computes the Casimir interaction energy.}, the remaining contribution becomes finite and thus may be identified with the Casimir interaction energy.

At the same time, the string radius is determined by the energy scale of that phase transition with the symmetry breaking, where strings were generated.
For strings on the Grand-Unified-Theory (GUT) scale,
 one has $r_0 \simeq 10^{-28}\,\text{cm}$. On these energy (or length)  scales the cone vertex is to be deformed into the smooth cap, which continuously transit to the external (with respect to the string core)
 conical domain. Therefore is gives rise to a question how the transverse string's size affects on the quantum-field effects in the string neighbourhood.

In the present work we consider the problem of vacuum interaction of cosmic strings if to take into account their finite width. The computation is carried out within the trln-formalism. In this so called global approach, one starts with that formal expression for the total vacuum energy, which is determined by the effective action.

We work with $ G=\hbar=c=1$ units, the metric signature is~$(+\, , -\, , -\,
,-\,)$, the definition of Riemann tensor is $R^{\mu}{}_{\nu\lambda\rho}=\Gamma^{\mu}_{\nu\lambda,\rho}-\pp$.

\section{Space-time of a system of parallel thick strings}

Let consider the four-dimensional space-time ($\mathcal{V}_4$) being the Cartesian product of the two-dimensional Minkowski space-time ($\mathbb{M}_{1,1}$, with coordinates $(t,z)$ on it), and two-dimensional Riemannian surface ($\mathcal{V}_2$, with coordinates $\mathbf{x}=(x,y)$).
Since any two-dimensional Riemannian surface is locally-conformal to the Euclidean plane, it allows to bring the metric on $\mathcal{V}_4$ to the form
\begin{align}\label{s0}
ds^2 = d t^2 - d z^2 - e^{- \sigma
(\mathbf{x})}\left(d x^2 + d y^2\right)
\, .
\end{align}
Let specify $\sigma
(\mathbf{x})$ as
\begin{align}\label{s1}
  \sigma
(\mathbf{x})=\suma\sigma_a( |\mathbf{x}-\mathbf{x}_a|)\,,
\end{align}
where $|\cdot|$ stands for the Euclidean norm:
$ |\mathbf{x}|:= (x^2 + y^2)^{1/2}$, while $\mathbf{x}_a$ is a location of the center of $a$-th string core.

The Ricci scalar of the metrics (\ref{s0}) equals
\begin{align}
{\rm R}= \e^\sigma\suma \el\sigma_a\, ,\nn
\end{align}
where $\el$ stands for the two-dimensional Euclidean laplacian. Thus the curvature may be presented  in additive form
\begin{align}
{\rm R}=  \suma {\rm R}_a\, .\nn
\end{align}

If   supports $\Omega_a$ of partial contributions  $\el\sigma_a$ are compact and do not intersect each other, then we deal with the ultra-static spacetime. Fixing $t$ and $z$, on each two-dimensional plane  $(xy)$ the curvature does not vanish in a number of domains $\Omega_a$.

In order to satisfy this, the partial  $\sigma_a$ should satisfy the two-dimensional Laplace equation outside the string cores:
\begin{align}\label{e4}
\sigma_a(\mathbf{x})=\left\{
\begin{array}{ll}
\sigma_a^{\ie}=2(1-\beta_a)f_a\big(|\mathbf{x}-\mathbf{x}_a| \big)\,,& \quad| \mathbf{x}-\mathbf{x}_a|\leqslant r_a\,;\\[2pt]
\sigma_a^{\et}=2(1-\beta_a)\ds\ln\frac{|\mathbf{x}-\mathbf{x}_a|}{r_a} \,, & \quad| \mathbf{x}-\mathbf{x}_a|\geqslant r_a\,,
\end{array}\right.
\end{align}
where all parameters $\beta$ are assumed to be \mbox{$\beta_a<1$}\,, and $f_a(\rho_a)$ is a twice differentiable function (of argument $\rho_a:=| \mathbf{x}-\mathbf{x}_a|$), which satisfies the boundary conditions
\begin{align}\label{e5}
f_a(r_a)=0\,,\qquad f_a'(r_a)=\frac{1}{r_a}\,.
\end{align}

With such a choice of the conformal factor, the scalar curvature vanishes everywhere where $| \mathbf{x}-\mathbf{x}_a| > r_a$ (for all $a$), and on this domain the metrics coincides with that one of a system of parallel infinitely-thin cosmic strings \cite{Letelier}. The criterion of the possibility to compute effects within the perturbation-theory framework (and to realize the desired smallness of perturbations) is got by the smallness of parameters  $$\bs_a := 1-\beta_a \,,$$
which are nothing but  complements to each $\beta_a$.
It is assumed that for the  GUT-strings $\bs$ has order $\sim 10^{-6}$.

Therefore, the space-time with metric (\ref{s0}) and the conformal factor (\ref{e4}) is to be considered  as a space-time generated by the system of $N$ parallel cosmic strings with non-zero width, with the scalar curvature
\begin{align}
\rm R(\mathbf{x})=\left\{
\begin{array}{ll}
{\rm R^{\ie }(\mathbf{x})}=\Delta\kn\sigma_a^{\ie }=e^{\sigma}\,\el \sigma_a^{\ie }\, &\qquad |\mathbf{x}-\mathbf{x}_a|\leqslant r_a\,;\\
{\rm R^{\et}}(\mathbf{x})=0 &\qquad |\mathbf{x}-\mathbf{x}_a| > r_a \quad \text{for all } a=1,2, \pp, N\,.
\end{array}\right. \nn
\end{align}
Such a metric is a solution of the Einstein equations with source ${\rm T}_{\mu\nu} $, the energy density of which reads
\begin{align}
{\rm T}_t^t=\frac{\rm R}{16\pi}=\frac{ \e^{\sigma}}{16\pi}\,\suma \el \sigma_a \,.\nn
\end{align}
Hence, the energy of a thick string equals
\begin{align}
\int {\rm T}^t_t  \sqrt{-g}\,dz\,d^2 x=\frac{1}{16\pi}\int\limits_{-\infty}^{\infty}dz\,\suma\int d\mathbf{x}\,\el\sigma_a^\ie =
\int\limits_{-\infty}^{\infty}dz\,\suma\frac{1-\beta_a}{8\pi}\int d\mathbf{x}\,\el\,f_a\, .\nn
\end{align}
Therefore the quantity
\begin{align} \label{e7}
\mu_a:=\frac{1-\beta_a}{8\pi}\int d\mathbf{x}\,\el f_a\,
\end{align}
is to be regarded as the energy-per-unit-length of $a$-th string.

The consideration of infinitely thin sting assumes the limit $r_a\rightarrow 0^+$, where the support $\Omega_a$ tends to a single point $\mathbf{x}_a$, with fixed value of integrals over $\Omega_a$. It corresponds to the fixation of the string's linear density. In this limit the exponent  (\ref{s0}) goes to
\begin{align}\label{inf-thin}
\sigma(\mathbf{x})=2\suma  (1-\beta_a)\ln|\mathbf{x}-\mathbf{x}_a|\, .
\end{align}


Acting by laplacian on it and regarding the limit in sense of distributions (see, e.g. \cite{Gelfand}), we get
\begin{align}\label{lim}
 \lim\limits_{r_a\rightarrow 0^+} \el \sigma_a=4\pi(1-\beta_a)\,\delta^{2}(\mathbf{x}-\mathbf{x}_a)\, .
 \end{align}
 Therefore, in addition to (\ref{e5}), for the functions  $f_a$ we should require the normalization
\begin{align}
\int d\mathbf{x}\,\el f_a=2\pi\, .\nn
 \end{align}

Then from eq.\,(\ref{e7}) we infer
\begin{align}
\mu_a=\frac{1-\beta_a}{4}\, , \nn
 \end{align}
and thus in the limit $r_a\rightarrow 0^+$ the following heuristic expression holds:
\begin{align}\label{y1}
{\rm T}_{tt} (\mathbf{x}) =\e^{\sigma(\mathbf{x})}\,\suma \mu_a\,\delta^{2}(\mathbf{x}-\mathbf{x}_a)\,.
\end{align}

If $a$ takes the single value and $r_1 \to 0^+$, the metric (\ref{s0}) is the one of an infinitely-thin cosmic string developed in \cite{Vilenkin2}.
Later it was shown in \cite{Letelier}, that the corresponding solution with conformal factor (\ref{inf-thin}) does represent the metric of $N$ parallel infinitely-thin cosmic strings with source (\ref{y1}). The two-dimensional surface $(xy)$ represents the locally-flat hypersurface (of the spatial subspace with fixed time) with a number of conical singularities located at  $\mathbf{x}=\mathbf{x}_a$, while the parameter $\mu_a$ defines the
angular deficit  $\delta{\varphi}_a=8\pi\,\mu_a=2\pi\beta^{\kn\prime}_a$, related with $a$-th singularity.

Hereafter we shall assume that the conformal coordinates map the surface $(xy)$ globally. For a single singularity it takes place if $\mu<1/4$, while for ($N\geqslant 2$) if   $$\sum_{a=1}^{N} \mu_a< \frac{1}{2} \,;$$ so that the conical singularity does not acquire  the topology of a sphere  \cite{Deser,Deser1,Deser2,Deser3}.

In the  case of single infinitely-thin string ($N=1$) the space-time metric has two striking features: (i) the absence of any lengthy parameters and (ii) higher symmetry.
The first allows to state that for the massless field the vacuum expectation value of the energy-momentum tensor depends upon the distance ($r$) from the observation point to the singularity.  In four dimensions of a spacetime it scales as $\langle \kn T_{\mu\nu}\rangle_{\rm vac}\sim r^{-4}$. The second feature allows to separate variables in the field equation, to construct the Green's function analytically and to compute the renormalized  $\langle\kn T_{\mu\nu}\rangle_{\rm vac}$. In the case of two higher number of strings ($N\geqslant 2$)  the problem becomes too complicated, and the perturbation theory becomes of particular significance
\cite{Grats1995,Bordag2014,Grats2016}.

The consideration of strings with finite diameter makes the problem even more complicated technically, and in addition, requires the knowledge of the string-substance distribution inside the core. In other words, we need a concretization of the expression for  $\sigma_a^{\ie }$. A possible way to smoothing the cone  vertices is to specify the functions $\sigma_a^{\ie }$  in the form
\begin{align}\label{e10}
\sigma_a^{\ie}(\mathbf{x})=-(1-\beta_a)\bigg[1-\Big(\frac{\mathbf{x}-\mathbf{x}_a}{r_a}\Big)^{\!2}\bigg]\, ,
\end{align}
what corresponds to the  so called ''ballpoint pen'' model known in the literature.

\begin{figure}[t]
 \begin{center}
 \includegraphics[width=9cm]{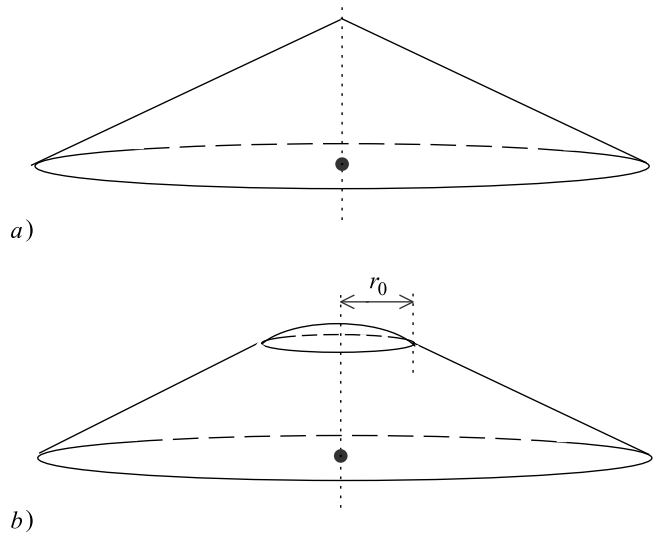} 
   \caption{Schematic illustration of two-dimensional projections of two models of cosmic-string space time: {\it a})  a cone which corresponds to the infinitely-thin cosmic string; {\it b}) so called ''ballpoint pen'' model, which corresponds to the cosmic string with radius $r_0$ and finite curvature inside.} \label{pen}
 \end{center}
\end{figure}

Therefore  the scalar curvature becomes
\begin{align}
\rm R(\mathbf{x})=\left\{
\begin{array}{ll}
{\rm R}^{\ie }_a(\mathbf{x})=4e^\sigma \beta^{\prime}_a/r_a^2\,, &\quad |\mathbf{x}-\mathbf{x}_a|\leqslant r_a\,;\\
{\rm R}^{\et}_a(\mathbf{x})=0\,, &\quad  |\mathbf{x}-\mathbf{x}_a| > r_a \quad \text{for all }a\, .
\end{array}\right.
\end{align}
For a single cosmic string the {\it schematic} illustration of conical singularity is presented on the Fig.\,\ref{pen}.

\section{Casimir energy in the trace-logarithm formalism}

For the real massless scalar field $\phi$, the action can be presented in equivalent form
\begin{align}\label{action}
S_{\rm \phi}=-\frac{1}{2}\int d^{\kn d}\akn x\,
\phi(x)\,L(x,
\partial)\,\phi(x)\,,
\end{align}
where $L(x, \partial)= \sqrt{-g} \,\Box$ is a whole field operator and $ \Box=\nabla_{\akn\mu}\!\nabla^{\mu}$ is a covariant Laplace\,--\,Beltrami operator on $\mathcal{V}_4$.

When the external conditions (the metric, boundaries, external fields, etc.) do  not depend upon time, the effective action $W_{\rm eff}$ is proportional to the total vacuum energy
${\cal E}_{\rm vac}$, namely:
$$W_{\rm eff}=-T {\cal E}_{\rm vac}\, ,$$
where $T$ stands for the total (infinite) time \cite{Peskin}.

At the other hand, in the trln-formalism
$$W_{\rm eff}=\frac{i}{2}\,\textmd{tr}\,\ln L=\frac{i}{2}\,\ln \det L\, $$
and hence, the vacuum energy, defined via
the effective action, reads
\begin{align}\label{Evac}
{\cal E}_{\rm vac}=-\frac{i}{2T}\,\ln \det L\, .
\end{align}

Now represent $L(x,
\partial)$ as
\begin{align}\label{K1}  L(x,
\partial)=\partial^2+ \delta L(x, \partial)\,,
\end{align}
where the perturbed operator is given by
\begin{align}
 \delta L(x,
\partial)=\sqrt{-g}\,\square -
\partial^2
\end{align}
and $
\partial^2:=\partial_t^2 - \partial_z^2 - \partial_x^2 -  \partial_y^2$ is a flat (Minkowskian) dalembertian.

In our case with metric (\ref{s1}), the perturbation operator
$\delta L(x, \partial)$ takes the form
\begin{align}\label{L2}
\delta L(x, \partial)=
\Lambda(\mathbf{x})\left(\partial^2_t - \partial^2_z\right)\,
,\qquad
\Lambda(\mathbf{x})=\e^{-\sigma(\mathbf{x})}-1 \, .
\end{align}
Considering $\delta L$ as perturbation, we get formal expansion
\begin{align}\label{L3}
\ln\det L&=\ln\det\left(\partial^2 +
\delta L\right)=  \nn \\
&=\ln\det\left(\partial^2\right)+
\ln\det\left(1+\partial^{-2}\,\delta L\right)= \nn \\
&=\mathrm{tr}\ln\left(\partial^2\right)+
\mathrm{tr}\ln\left(1+\partial^{-2}\,\delta L\right)=\nn \\
&=\mathrm{tr}\ln(\partial^2) +
\mathrm{tr}\left(\partial^{-2}\,\delta L\right)
-\frac{1}{2}\,\mathrm{tr}\left(\partial^{-2}\delta L\,\partial^{-2}\delta L\right)
+\pp\, .
\end{align}
As it will be demonstrated below, the first two terms do not contribute, hence in our framework with  the declared accuracy  we  infer
\begin{align}\label{Weff}
W_{\rm eff}=-\frac{i}{4}\,\mathrm{tr} \left(\partial^{-2}\delta L\,\partial^{-2}\delta L\right),
\end{align}
plus terms with higher powers of $\bs$.

The expression (\ref{Weff}) is well-defined if all operators in it (and their products) belong to the operators-with-trace class
(see, e.g., \cite{Reed}). Otherwise, the trace represents some  formal expression which diverges. Our goal is to take advances of the dimensional-regularization technique.
  However,  it gives rise to another problem proper to the curved background. As it was shown by Hawking,
\cite{Hawking}, in this case there is no natural recept which dimensions should be specified for the analytical continuation. The result may depend drastically upon the choice and may differ results got by another regularization techniques.  The way proposed in
\cite{Hawking} for four-dimensional curved space-time, consists in the construction of a direct (Cartesian) product of the curved $\mathcal{V}_4$ space-time under consideration and fictitious $(D-4)$-dimensional flat space. It is demonstrated that the analytical continuation within the proposed way coincides with the result obtained by the method of generalized
$\zeta$~-function.

In our case under interest, the space-time $\mathcal{V}_4$ {\it already} represents the product of curved  $\mathcal{V}_2$ space and Minkowski $\mathbb{M}_{1,1}$, and thus the Hawking's prescript works {\it a priori}, so the regularization will be applied to the total dimensionality, fixing the dimension of curved subspace.

For the massless field, the first two terms in  (\ref{L3}) contain single Green's function  $\partial^{-2}$ inside themselves and therefore correspond to the ''tadpole'' diagrams.
In the dimensional-regularization framework, these diagrams are regarded as yielding  zero  contribution. The motivation of it has physical \cite{Itzyk} and mathematical \cite{Gelfand} grounds and described in the popular textbooks.

 Therefore for the Casimir contribution to the total vacuum energy (which diverges!), to the lowest order of the perturbation theory, we have to restrict our consideration by the third term in the final expansion (\ref{L3}).
As result, for the vacuum energy  (\ref{Evac}) we get eventually:
\begin{align}\label{vac2}
{\cal E}_{\rm vac}=\frac{i}{4T}\,\textmd{tr}\left(\partial^{-2}\,\delta L\,\,
\partial^{-2}\, \delta L\right)  .
\end{align}
It allows to use the standard Fourier-basis formalism. It yields:
  \begin{align}\label{int3}
{\cal E}_{\rm vac}=\frac{i}{4T}\int \frac{d^{\kn 4}\akn
k}{(2\,\pi)^4}\,\frac{d^{\kn 4}\akn p}{(2\pi)^4}\,\,\frac{\delta L\left(k, i(p+k)\right)\,
\delta L\left(-k, ip\right)}{p^2 (p+k)^2}\ ,
\end{align}
where
\begin{equation}\label{deltaL1}
\delta L(k, ip):=\int d^{\kn 4}\akn  x\, e^{ikx}\!\left[\left.
\delta L(x,
\partial)\right|_{\partial\to-ip}\right] .
\end{equation}
In our problem we have from (\ref{L2}):
\begin{align}
\delta L(k,
ip)=-\Lambda(k)\left(p_{{\kern 0.5pt}0}^2 - p_z^2\right)\kn,
\end{align}
and thus one obtains
\begin{align}\label{Cas}
{\cal E}_{\rm vac}=\frac{i}{4T}\int \frac{d^{\kn 4}\akn
k}{(2\,\pi)^4}\,\frac{d^{\kn 4}\akn p}{(2\pi)^4}\,\frac{(p_{{\kern 0.5pt}0}^2-p_z^2)^2}{p^2\,
(p+k)^2}\,\Lambda(k)\,\Lambda(-k)\, .
\end{align}

The integral over  $d^{\kn 4}\akn p$ in (\ref{Cas}) diverges, but it has the form which is standard for the dimensional-regularization technique.
The Wick rotation,
$$ p^0=i\,p^0_{\eu}\, ,\quad d^{\kn 4}\akn p =i\,d^{\kn 4}\akn p_{{\kern 0.5pt}\eu}\, ,\quad p^2=-p_{\eu}^2\, ,$$
and subsequent replacement $d^{\kn 4}\akn p$ by $d^{\kn D}\akn  p_{{\kern 0.5pt}\eu}$, $ D=(4-2\varepsilon)$ bring the integral
 over $d^{\kn 4}\akn p$ into the  form \cite{Itzyk}
$$
i\,{\lambda}^{2\varepsilon}\int \frac{d^{\kn D}\akn
p_{{\kern 0.5pt}\eu}}{(2\pi)^{D}}\,\frac{\left(p_0^2 -
p_z^2\right)^2}{p^2_{\eu} (p+ {k})^2_{\eu}}
=i\,\frac{{\lambda}^{2\varepsilon}}{(4\pi)^{D/2}}\,\frac{2\,D}{(D-2)}\,
\frac{\Gamma^2{\kern -1pt}(D/2)}
{\Gamma(D+2)}\, \Gamma\Big(\frac{4-D}{2}\Big)\,\left(k_{\eu}^2\right)^{D/2}\!. $$
Here the parameter $\lambda$ (with dimensionality of length) is introduced in order to keep the dimension of the whole expression under regularization.

The subsequent integration over $dk^0 \,dk^z$ is a bit tricky, since the integrand contains square of
$$\Lambda(k)=4\pi^2 \delta(k^0)\,\delta(k^z)  \,\Lambda(\mathbf{k})\,,$$
where $\Lambda(\mathbf{k})$ is a Fourier-transform of $\Lambda(\mathbf{x})$ given by (\ref{L2}), with suggestive notation for $\mathbf{k}$. The problem is to be resolved in the standard for QFT way:
the first delta-function with accompanying measure will yield  unity when integrated, and will set the argument of the second delta to zero. Then we represent
$$ \delta(k^0) \Big|_{k^0=0}=\frac{1}{2\pi}\int \e^{i  k^0 t}\,dt \Big|_{k^0=0} =\frac{1}{2\pi}\int dt =\frac{T}{2\pi}\,,$$
where $T$ stands for the total (infinite) time. The same argumentation for $k^z$ yields
$$ \delta(k^z) \Big|_{k^z=0}=\frac{1}{2\pi}\int \e^{-i  k^z z}\,dz \Big|_{k^z=0} =\frac{1}{2\pi}\int dz =\frac{Z}{2\pi}\,,$$
where $ Z$ stands for the total (infinite) string length. Therefore the remaining integration is two-dimensional one over $d\mathbf{k}$. By the same reasons mentioned above, $k_{\eu}^2$ becomes $\mathbf{k}^2$.

Now the regularized ${\cal E}_{\rm vac}$ in eq.\,(\ref{Cas}) becomes
\begin{align}\label{int6}
{\cal E}_{\rm vac}^{\rm reg}=-Z\,\frac{{\lambda}^{2\varepsilon}}{2(4\pi)^{D/2}}\,
\frac{D}{(D-2)}\,
\frac{\Gamma^2{\kern -1.3pt} (D/2 )}
{\Gamma(D+2)}\, \Gamma\Big(\frac{4-D}{2}\Big)\,\int
\frac{d \mathbf{k}}{(2\pi)^2}\,
\left({\mathbf{k}}^2\right)^{D/2}\Lambda(\mathbf{k})\,\Lambda(-\mathbf{k})\, .
\end{align}

The pre-factor of integral in
(\ref{int6}) has a simple pole at $D=4$, hence after the regularization removal the possible divergence can arise due to this pole, or due to the value of integral, or due to both reasons.

Now expand $|\mathbf{k}|^{D}=|\mathbf{k}|^{4-2\varepsilon}$ in small $\varepsilon$:
\begin{align}
|\mathbf{k}|^{4-2\varepsilon}=|\mathbf{k}|^{4}\big(1-2\varepsilon\ln|\mathbf{k}|\big)+{\cal{O}}(\varepsilon^2)\, .
\end{align}
Thus ${\cal E}_{\rm vac}^{\rm reg}$ rewrites as
\begin{align}\label{e13}
{\cal E}_{\rm vac}^{\rm reg}=-Z\,\frac{{\lambda}^{2\varepsilon}}{2(4\pi)^{2-\varepsilon}}\,
\frac{4-2\varepsilon}{(2-2\varepsilon)}\,
\frac{\Gamma^2\akn\left(2-\varepsilon\right)}
{\Gamma(6-2\varepsilon)}\, \Gamma\left(\varepsilon\right)\,\int
\frac{d \mathbf{k}}{(2\pi)^2}\,
|\mathbf{k}|^{4}\Big[\big(1-2\varepsilon\ln|\mathbf{k}|\big)+{\cal{O}}(\varepsilon^2)\Big]
\,\Lambda(\mathbf{k})\,\Lambda(-\mathbf{k})\, .
\end{align}
Now we encounter the following two-dimensional Fourier integrals:
\begin{align}\label{e13s}
I_1:=\int\frac{d \mathbf{k}}{(2\pi)^2}\,
|\mathbf{k}|^4\,\Lambda(\mathbf{k})\,\Lambda(-\mathbf{k})\,, \qquad\qquad I_2:=\int\frac{d \mathbf{k}}{(2\pi)^2}\,
|\mathbf{k}|^4\,\ln|\mathbf{k}|\,\Lambda(\mathbf{k})\,\Lambda(-\mathbf{k})\,.
\end{align}
The first one can be converted to the $\mathbf{x}$-integral
:
\begin{align}\label{e14}
I_1=\int d\mathbf{x}\big[\el\Lambda(\mathbf{x})\big]^2.
\end{align}
Computing the laplacian in polar coordinates, we should neglect $[\sigma'(\varrho)]^2$ with respect to $\sigma''(\varrho)$ and $\sigma'(\varrho)/\varrho$, since it contains extra small factor $\bs$; thus
\begin{align}\label{e14a}
I_1=\int\,d\mathbf{x}\big[\el \sigma(\mathbf{x})\big]^2
=\int d\mathbf{x}\,\rm{R}^2(\mathbf{x})\,,
\end{align}
where, with our accuracy, we shall fix $\e^{\sigma}$ equal to unity.

The corresponding integral $I_2$ in the $\mathbf{x}$-representation differs  from $I_1$ by the inverse Fourier transform of a logarithm \cite{Gelfand}:
\begin{align}\label{e15}
I_2=
-\frac{1}{2\pi} \int\,d\mathbf{x}\,d\mathbf{x}'\,\frac{\el\Lambda(\mathbf{x})\,\, \Delta\smash{^{\smash{\prime}}}_{\!\!\eu}   \Lambda(\mathbf{x}')}
{|\mathbf{x}-\mathbf{x}'|^2}\,,
\end{align}
and with the required accuracy we have similar
\begin{align}\label{e15a}
I_2=
-\frac{1}{2\pi} \int\,d\mathbf{x}\,d\mathbf{x}'\,\frac{\rm{R}(\mathbf{x})\,\rm{R}(\mathbf{x}')}
{|\mathbf{x}-\mathbf{x}'|^2}\, .
\end{align}

The pole contribution into the effective action (and to ${\cal E}_{\rm vac}^{\rm reg}$, respectively), which corresponds to the first integral  in eq. (\ref{e13}), should be ignored within the procedure of renormalization of   the effective action. The reason is that with the required accuracy, it corresponds to $a_2$-similar term in the Schwinger\,--\,de\,Witt expansion \cite{Birrell}. Notice, in the massless-field case, the $a_0$- and $a_1$-proportional terms of the expansion vanish completely.
In four dimensions of a space-time it provides the finiteness of the renormalized energy-momentum tensor, but does not guarantee the convergence of the remaining formal expression
for ${\cal E}_{\rm vac}^{\rm ren}$. The latter in our approximation takes the form
\begin{align}\label{e16}
{\cal E}_{\rm vac}^{\rm ren}=-\frac{\rm Z}{30\,(4\pi)^3}\int\,d\mathbf{x}\,d\mathbf{x}'\,\frac{\textrm{R}(\mathbf{x})\,\textrm{R}(\mathbf{x}')}
{|\mathbf{x}-\mathbf{x}'|^2}\,.
\end{align}
Within the problem formulation, common for the Casimir effect, the criterion of elicitation of the Casimir contribution from the total vacuum energy is a dependence upon the distance between ''walls'' (or other interacting objects). One proves that for the finite-sized bodies, separated by the finite distance, the corresponding Casimir contribution into the total (generally, diverging) vacuum energy turns out to be {\it finite}
(see, e.g. \cite{Bordag}). In our problem this condition holds, since we demand that the supports $\Omega_a$ of partial curvatures ${\rm R}_a$  do not intersect.
In particular,  this prescript allows  to neglect those terms in the integrand, which contain products
${\rm R}_a{\rm R}_a$.  Furthermore, with our accuracy $\e^{\sigma}=1+\mathcal{O}(\bs)$, and
 the  partial contributions ${\rm R}_a$ constitute the curvature ${\rm R}$  {\it additively}, and thus  to  the lowest in   $(\beta^{\prime}_a)$ order  the Casimir interaction looks as pairwise:
\begin{align}\label{e16a}
{\cal E}_{\rm vac}^{\rm ren}=-\frac{\rm Z}{15\,(4\pi)^3}\sum_{a<b}\int\,d\mathbf{x}\,d\mathbf{x}'\,\frac{\textrm{R}_a(\mathbf{x})\,\textrm{R}_b(\mathbf{x}')}
{|\mathbf{x}-\mathbf{x}'|^2}\,.
\end{align}
For two strings, separated by distance $d$, the Casimir energy
 (\ref{e16}) is expressed as
\begin{align}
{\cal E}_{\rm cas}=-\frac{Z}{15\,(4\pi)^3}\int d\mathbf{x}\,d\mathbf{x}'\,\frac{{\rm R_1^\ie }(\mathbf{x})\,{\rm{R_2^\ie }}(\mathbf{x}')}
{|\mathbf{x}-\mathbf{x}'+\mathbf{d}|^2}\, ,\qquad \mathbf{d}=\mathbf{x}_1 -\mathbf{x}_2 \, .
\end{align}

Introducing two polar coordinate systems with apex on the center of two strings, both two angular integrations may be carried out with the help of  table integral
\begin{align}
\int\limits_0^{2\pi}\frac{d\varphi}{A+B\cos\varphi}=\frac{2\pi}{\sqrt{A^2-B^2}}\,.\nn
\end{align}
It yields
\begin{align}
\frac{{\cal E}_{\rm cas}}{ Z}=\frac{16}{15\pi}\,\frac{\mu_1\mu_2}{r_1^2r_2^2} \int\limits_0^{r_1}\varrho\kn d\varrho\int\limits_0^{r_2}
\frac{\varrho' d\varrho'}{\sqrt{[(d+\varrho')^2-\varrho^2][(d-\varrho')^2-\varrho^2]}}\,.\nn
\end{align}

For simplicity, we take both string diameters equal:
$r_1=r_2=r_0$. Introducing   $\xi:=r_0/d$, the Casimir energy may be expressed as
\begin{align}\label{e18}
\frac{{\cal E}_{\rm cas}}{ Z}=\frac{4}{15\pi}\,\frac{\mu_1\mu_2}{d^2}\,\frac{1}{\xi^{4}}\int\limits_{0}^{\xi^2} d\eta \ln\frac{1+\eta-\xi^2+\sqrt{(1+\eta-\xi^2)^2-4\eta}}{2}\,,
\end{align}
where the requirement $\xi<1/2$ means that two strings do not intersect each other.

   \begin{figure}[t]
 \begin{center}
\includegraphics[width=10cm]{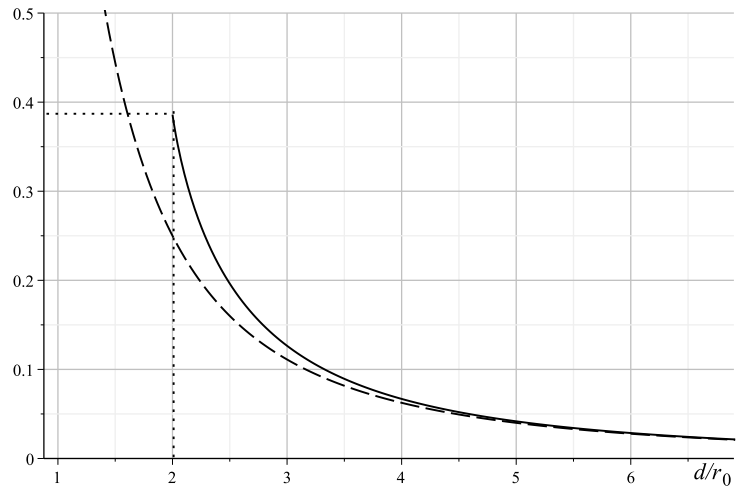}  \caption{Absolute dependence of the Casimir effect on the distance between centers  for fixed (and equal) string radii (in $r_0=1$ units, solid) compared with the infinitely-thin string  (dashed). } \label{d-dev_abs}
 \end{center}
\end{figure}

Finally, integrating over $\eta$, for the energy of Casimir interaction of two parallel thick cosmic strings per unit length  we arrive at
\begin{align}\label{e18f}
{\cal E}_{\rm cas}\Big|_{Z=1}=-\frac{4}{15\pi}\,\frac{\mu_1\mu_2}{d^2}\,\frac{1}{\xi^{2}}\bigg[ 1-2 \ln \frac{1+\sqrt{1-4\xi^2}}{2} -
\frac{1-\sqrt{1-4\xi^2}}{2\xi^2}  \bigg]\,.
\end{align}
The plot of the Casimir energy (omitting the common  pre-factor $(-4\mu_1
\mu_2/15 \pi)$) versus the inter-string distance is presented on Fig.\,\ref{d-dev_abs}.

For the case of ultra-thin strings $(d\gg r_0)$, we expand in $\xi\ll 1$:
\begin{align} {\cal E}_{\rm cas}\Big|_{Z=1}=-\frac{4}{15\pi}\,\frac{\mu_1
\mu_2}{d^2}  \bigg[ 1+\frac{r_0^2}{d^2} + \frac{5}{3}\frac{r_0^4}{d^4}+{\cal{O}}(d^{-6})\bigg]\, ,
\end{align}
what to the leading order coincides  with the results get for the infinitely-thin cosmic strings \cite{Grats1995, Bordag2014, Grats2016}.

\begin{figure}[h]
 \begin{center}
\includegraphics[width=11cm]{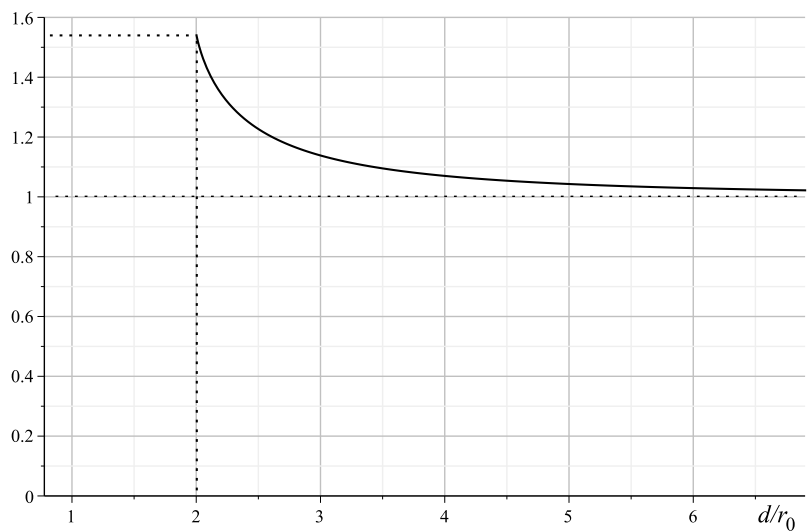}  \caption{Relative   Casimir effect energy versus the inter-string distance,
with respect to the infinitely-thin string.} \label{d-dev_rel}
 \end{center}
\end{figure}

The plot of the relative influence of the string width on the Casimir energy is presented on Fig.\,\ref{d-dev_rel}.

\section{Discussion}

In the present work we examined the influence of the transverse size of cosmic strings  on the character of their vacuum  interaction.
 Idealized model of the string's space-time, where the curvature  does not vanish on the cone's apex only and has a delta-like singularity there, is valid on the large distances from the cosmic  string. Meanwhile, the core radius is determined by the energy scale of the Universe's phase transition (with breaking symmetry) at the epoch where strings were generated. For the GUT-strings the radius is of order
 $r_0\sim 10^{-28}\text{cm}\,.$ At these lengthy scales, the cone apex should be considered not as a single point, but as smoothed cap, which is continuously united with the external conical region. At the other hand, the way of smoothing (the choice of smooth functions $f_a(\varrho)$ in eq.(\ref{e4})) should not  have a crucial significance if it admits the
 proper limit, corresponding to the infinitely-thin cosmic string.
 Our way is conventional and widely-used and, besides that it satisfies all the necessary requirements, it allows to carry out all computations we need, analytically.

 In the work we have used the method proposed in our work on the Casimir effect of infinitely-thin cosmic strings \cite{Grats1995}. It was continued in subsequent works \cite{Grats2016}
and was of usage for the research of other field effects on the conical backgrounds
 \cite{Grats2017,Grats2020}. The key point of the method is a usage of conformal coordinates on the two-dimensional submanifold transverse to a string.  With help of dimensional regularization of the total vacuum energy we separate the finite final expression for the Casimir energy.

Our computation shows that interactions in the cosmic-string net look as pairwise only in the lowest order in  $(1-\beta_a)$. It reflects that fact that the field theory (classical or quantum) is effectively non-local -- in that sense that solutions of the field equations carry the information on the space-time structure {\it in toto}.
In our case it appears
as follows: the perturbed operator $\delta L$ (\ref{L2}) depends upon the partial contributions from the single string not additively, but in multiplicative way (through the conformal-factor exponential in (\ref{s0})). The change of the string number in the net changes the value of ''pairwise'' contributions into the total vacuum energy.
If so, it changes the structure of interactions  {\it in toto}. However, the effects related with the finite width, are valuable at the distances of several core diameters.

The latter has a simple qualitative explication. Indeed, the $\Pi$-theorem, which is central for the dimensional analysis, allows to state that for single string and for the massless field the vacuum expectation value
scales as
$$\left\langle{\rm T}_{tt}\right\rangle_{\rm vac}^{\rm ren}\sim \frac{1}{r^4}\, \Theta\Big(\frac{r_0}{r}\Big)\, , $$
where $\rho$  is a distance from the observation point to the string's center, $ \Theta$ -- some monotonically decaying function, which goes like a constant  in the limit $r_0\rightarrow 0^+$. Hence one might expect that the influence of transverse size will be perceptible on the scales of several $r_0$\,. In analogy, for the Casimir effect the non-zero string width should make the significant influence on the same-order distances.  Our computation confirms these estimates quantitatively;  the resulting expression is in perfect agreement with the computations on the Casimir interaction between zero-width cosmic strings \cite{Grats1995,Bordag2014,Grats2016}.

\medskip

{\bf Acknowledgment.} The research  was carried out within the framework of the scientific program of the National Center
for Physics and Mathematics, the project ''Particle
Physics and Cosmology.''


\begin{thebibliography}{29}

\bibitem{Bordag}
M.\,Bordag, G.\,L.\,Klimchitskaya, U.\,Mohideen, V.\,M.\,Mostepanenko,
{\it Advances in the Casimir Effect} (Oxford University Press, Oxford,
2009)

\bibitem{Dowker}
 G.\,W.\,Gibbons, S.\,W.\,Hawking, T.\,Vachaspati (eds.), {\it The Formation and Evolution of Cosmic Strings} (Cambridge University Press,
Cambridge, 1990)

\bibitem{Vilenkin}
A.\,Vilenkin, E.\,P.\,S.\,Shellard, {\it Cosmic Strings and Other Topological
Defects} (Cambridge University Press, Cambridge, 1994)



\bibitem{Copeland}
E.\,J.\,Copeland, L.\,Pogasian, T.\,Vachaspaty, Class. Quantum Gravity
  \textbf{28}, 204009, (2011)

  \bibitem{Bordag1990}
M.\,Bordag, Ann. Phys. (Berlin) \textbf{47}, 93 (1990)

\bibitem{Bordag2014}
J.\,M.\,Mu\~{n}oz-Casta\~{n}eda, M.\,Bordag, Phys. Rev. D.  \textbf{89}, 6, 065034, (2014)

\bibitem{Grats2016}
Yu.V.\,Grats, Theor. Math. Phys. \textbf{186}, 205, (2016)



\bibitem{Grats1995}
D.\,V.\,Gal'tsov, Yu.V.\,Grats, A.\,B.\,Lavrent'ev, Phys. Atom. Nucl. \textbf{58},
516 (1995)

\bibitem{Frolov}
T.\,M.\,Helliwell, D.\,A.\,Konkowski, Phys. Rev. D \textbf{34}, 1918 (1986)

\bibitem{Frolov1}
 V.\,P.\,Frolov, E.\,M.\,Serebriany, Phys. Rev. D \textbf{35}, 3779 (1987)

\bibitem{Frolov2}
 B.\,Linet, Phys. Rev. D. \textbf{35}, 536, (1987)

 \bibitem{Frolov3}
J.\,S.\,Dowker, Phys. Rev. D \textbf{36}, 3095 (1987)

 \bibitem{Frolov4}
 J.\,S.\,Dowker, Phys. Rev. D \textbf{36}, 3742 (1987)

\bibitem{Frolov5}
 W.\,A.\,Hiscock, Phys. Lett. B  \textbf{188}, 317, (1987)



\bibitem{Fursaev}
D.\,V.\,Fursaev,  Class. Quantum Gravity  \textbf{11}, 1431, (1994)

\bibitem{Letelier}
 P.\,C.\,Letelier, Class. Quantum Gravity \textbf{4}, 75 (1987)

\bibitem{Gelfand}
 I.\,M.\,Gel'fand and G.\,E.\,Shilov, {\it Generalized Functions: Properties and operations} (Academic Press, Waltham, 1964)


 \bibitem{Vilenkin2}
 A.\,Vilenkin, Phys. Rev. D \textbf{23}, 852 (1981)





\bibitem{Deser}
 S.\,Deser, R.\,Jackiw, G.\,'t Hooft,   Ann. of Phys. (USA), \textbf{152},
220, (1984)

\bibitem{Deser1}
 J.\,R.\,Gott, M.\,Alpert, Gen. Relativ. Gravit. \textbf{16}, 243 (1984)

\bibitem{Deser2}
 S.\,Giddings,  J.\,Abbott, K.\,Kuchnar,   Gen. Relat. Gravit. \textbf{16}, 751, (1984)

\bibitem{Deser3}
R.\,Jackiw, Nucl. Phys. B  \textbf{252}, 343, (1985)









\bibitem{Peskin}
M.\,E.\,Peskin, D.\,V.\,Schroeder, {\it An Introduction to Quantum Field Theory} (Addison-Wesley, Reading, 1995)

\bibitem{Reed}
M.\,Reed, B.\,Simon, {\it Methods of mathematical physics III: Scattering theory} (Academic
Press,  New York, 1979)

\bibitem{Hawking}
S.\,W.\,Hawking, Commun. Math.  Phys.  \textbf{55}, 133, (1977)

 \bibitem{Itzyk}
 C.\,Itzykson, J.\,B.\,Zuber, {\it Quantum Field Theory} (McGraw-Hill, 1980)



\bibitem{Birrell}
N.\,D.\,Birrell and P.\,C.\,W.\,Davies, {\it Quantum Fields in Curved Space} (Cambridge University Press, Cambridge, 1982)

\bibitem{Grats2017}
Y.\,V.\,Grats, P.\,Spirin, Eur. Phys. J. C \textbf{77}, 101, (2017)

\bibitem{Grats2020}
Y.\,V.\,Grats, P.\,Spirin, Int.\,J.\,Mod.\,Phys.\,A,  \textbf{35}, no.02n03, 2040030 (2020).

\end{thebibliography}
\end {document}